\documentclass[fleqn,usenatbib,twocolumn]{mnras}

\usepackage{newtxtext,newtxmath}

\usepackage[T1]{fontenc}
\usepackage{ae,aecompl}

\usepackage{graphicx}	
\usepackage{amsmath}	
\usepackage{hyperref}
\usepackage[utf8]{inputenc}
\usepackage{mathrsfs}
\usepackage{bm}
\usepackage{hyperref}
\usepackage[dvipsnames]{xcolor}
\usepackage{booktabs}

\newcommand{\be}{\begin{equation}}	
\newcommand{\ee}{\end{equation}}	
		
\newcommand{\refig}[1]{Fig.~\ref{#1}}
\newcommand{\refeq}[1]{Eq.~\ref{#1}}
\newcommand{\dd}{{\rm d}}

\title[MRI at very high magnetic Prandtl numbers]
{MRI-driven dynamo at very high magnetic Prandtl numbers}
  
  \author[Guilet et al.]{J\'er\^ome Guilet$^{1}$, Alexis Reboul-Salze$^{1,2}$, Rapha\"el Raynaud$^3$, Matteo Bugli$^1$, Basile Gallet$^4$\\
$^1$ Université Paris-Saclay, Université Paris Cité, CEA, CNRS, AIM, 91191, Gif-sur-Yvette, France \\ 
$^2$ Max Planck Institute for Gravitational Physics (Albert Einstein Institute), D-14476 Potsdam, Germany \\
$^3$Université Paris Cité, Université Paris-Saclay, CEA, CNRS, AIM, F-91191, Gif-sur-Yvette, France\\
$^4$ Universit\'e Paris-Saclay, CNRS, CEA, Service de Physique de l’Etat Condens\'e, 91191, Gif-sur-Yvette, France.}

\date{Accepted XXX. Received YYY; in original form ZZZ}

\pubyear{2015}

\begin{document}
\label{firstpage}
\pagerange{\pageref{firstpage}--\pageref{lastpage}}
\maketitle

\begin{abstract}
The dynamo driven by the magnetorotational instability (MRI) is believed to play an important role in the dynamics of accretion discs and may also explain the origin of the extreme magnetic fields present in magnetars. Its saturation level is an important open question known to be particularly sensitive to the diffusive processes through the magnetic Prandtl number Pm (the ratio of viscosity to resistivity). Despite its relevance to proto-neutron stars and neutron star merger remnants, the numerically challenging regime of high Pm is still largely unknown. Using zero-net flux shearing box simulations in the incompressible approximation, we studied MRI-driven dynamos at unprecedentedly high values of Pm reaching 256. The simulations show that the stress and turbulent energies are proportional to Pm up to moderately high values ($\mathrm{Pm} \sim 50$). At higher Pm, they transition to a new regime consistent with a plateau independent of Pm for $\rm Pm \gtrsim 100$. This trend is independent of the Reynolds number, which may suggest an asymptotic regime where the energy injection and dissipation are independent of the diffusive processes. Interestingly, large values of Pm not only lead to intense small-scale magnetic fields but also to a more efficient dynamo at the largest scales of the box.
\end{abstract}

\begin{keywords}
stars: magnetars -- neutron star mergers -- supernovae: general --  MHD -- instabilities -- magnetic fields
\end{keywords}

\section{Introduction}
The magnetorotational instability (MRI) is believed to play a crucial role in a large number of astrophysical objects. This includes accretion discs around a variety of objects \citep{balbus98},  neutron star mergers \citep[e.g.][]{siegel13,kiuchi14,kiuchi18,guilet17}, stellar mergers \citep[e.g.][]{schneider19} and core-collapse supernovae \citep[e.g.][]{akiyama03,obergaulinger09,masada07,guilet15b,moesta15,reboul-salze21}. In neutron star mergers and core-collapse supernovae, the MRI may generate extreme magnetic fields reaching up to $10^{15}-10^{16}\,\mathrm{G}$, potentially leading to the formation of a magnetar \citep{reboul-salze21,reboul-salze21b}. In combination with fast rotation, such extreme magnetic fields can trigger powerful magnetorotational explosions\footnote{Note that with \emph{magnetorotational explosion} we refer to the physical explosion mechanism of massive magnetised stars in rotation, not to be confounded with the MRI.} \citep[e.g.][]{takiwaki09,Kuroda20,bugli20,bugli21}. This so-called millisecond magnetar scenario may provide an explanation for outstanding explosions such as long gamma-ray bursts, hypernovae and superluminous supernovae \citep[e.g.][]{woosley10,metzger11}. The formation of a magnetar after a neutron star merger has also been proposed as a central engine of short gamma-ray bursts \citep[e.g.][]{moesta20}. 

The origin of the extreme magnetic fields of magnetars is still an important open question as several scenarios have been proposed, such as a convective dynamo in a fast rotating proto-neutron star (PNS) \citep{thompson93,raynaud20,raynaud22}, the Tayler-Spruit dynamo following fallback (Barrere et al, in prep), amplification in main sequence stellar mergers \citep{schneider19} or the fossil field scenario \citep{ferrario06}. To shed light on this question, it is important to assess the efficiency of each of the dynamo mechanisms in the conditions specific to a PNS. The impact of different ingredients on the MRI has been studied in recent years: the shear parameter \citep{masada12}, neutrino viscosity and drag \citep{guilet15}, stable stratification \citep{guilet15b,reboul-salze21b} and spherical geometry \citep{reboul-salze21}. A remaining open question is the dependence on diffusion coefficients. 
Local disc simulations have shown that the efficiency of the MRI is strongly correlated to the magnetic Prandtl number $\mathrm{Pm}$ (the ratio of viscosity to resistivity) when $\mathrm{Pm}\sim 0.1-16$ \citep{fromang07b,lesur07,simon09,longaretti10,shi16,potter17} and plateaus at $\mathrm{Pm}<0.1$ with an imposed external magnetic field \citep{meheut15}. Low and high Pm regimes are computationally challenging because a high numerical resolution is needed to resolve the small viscous (low Pm) or resistive (high Pm) scale. The regime of low Pm has attracted particular attention because it is relevant to most regions of accretion discs as well as to liquid metals (laboratory experiments, Earth core\dots), but the opposite regime of large $\mathrm{Pm}$ has not been targeted  specifically by previous studies of the MRI. This regime is relevant to PNSs \citep{thompson93,guilet15,lander21}, neutron star merger remnants \citep{rossi08}, which exhibit physical conditions similar to proto-neutron stars \citep{guilet17}, interstellar and intergalactic media \citep{schekochihin04}, and to the inner parts of some accretion discs \citep{balbus08,potter14,potter17}. In the latter case, the Pm-dependence of the MRI may drive an instability leading to variability in the accretion rate \citep{potter14,potter17,kawanaka19}.

Simulations of the MRI in a PNS including explicit diffusion coefficients are able to describe a realistic parameter regime for the (large) viscosity and thermal diffusion induced by neutrinos \citep{guilet15,reboul-salze21,reboul-salze21b}. By contrast, the physical resistivity in a PNS is much smaller than the values that can be afforded in these simulations. This translates into a large discrepancy between the huge values of $\mathrm{Pm} \sim10^{13}$ relevant to a PNS and the modest values $\rm Pm=4-16$ of these simulations. The regime of high magnetic Prandtl numbers is very challenging for numerical simulations because a high resolution is required to resolve the resistive length scale. Local models are best suited to tackle this problem owing to their simplicity and comparatively low computational cost. We probe unprecedentedly large $\mathrm{Pm}$ values (up to 256), using a zero-net flux shearing-box model inspired by a comparison between local and spherical models of the MRI \citep{reboul-salze21}. 


\section{Numerical setup}
\label{sec:setup}

Our simulations are designed to represent a
small region in the equatorial plane of a fast rotating PNS. The local dynamics is described in the framework of a Cartesian shearing box \citep[e.g.][]{goldreich65}. The coordinates $x$, $y$, and $z$ represent the radial, azimuthal and vertical directions, respectively, and the corresponding unit vectors
are $\bm{e_x}$, $\bm{e_y}$, and $\bm{e_z}$. The angular frequency
vector points in the $z$ direction $\bm{\Omega}=\Omega\, \bm{e_z}$,
while gravity and shear are in the $x$ direction. 
Neutrinos are assumed to be in the diffusive regime such that their effects on the dynamics can be described by a viscosity \citep{guilet15}. We used the incompressible approximation for the following reasons. Soundproof approximations are well justified for the study of the MRI in a PNS because the fluid velocity and the Alfv\'en speed are small compared to the sound speed \citep[$v/c_s<v_A/c_s \lesssim 10^{-2}$, see e.g.][]{guilet15,reboul-salze21,reboul-salze21b}. Limits to the applicability of the incompressible approximation in a PNS come mostly from the density gradient and the buoyancy driven by entropy and composition gradients. Neglecting the density gradient is an essential part of the local approximation, which is necessary in order to reach high magnetic Prandl numbers. On the other hand, while buoyancy could have been included with the Boussinesq approximation, it was shown by \citet{reboul-salze21b} to have relatively minor effects on the MRI in a PNS except for a small region near the equator. For this reason and for the sake of simplicity we adopted the incompressible approximation. The incompressible MHD equations in the shearing box approximation read
\begin{align}
  \label{eq:base1}
  \begin{split}
  \partial_t \bm{v}+\bm{v}\cdot \bm{\nabla} \bm{v} 
   &= - \frac{1}{\rho_0}\bm{\nabla}\Pi
       + \frac{1}{\mu_0 \rho_0}(\bm{\nabla}\times \bm{B})\times\bm{B}\\
       &\quad  + q\Omega x\partial_y \bm{v} + q\Omega v_x \bm{e_y} -2\bm{\Omega}\times\bm{v}
       + \nu\bm{\Delta}\bm{v}, 
    \end{split}\\
  \label{eq:base2}
  \partial_t \bm{B} &= \bm{\nabla}\times(\bm{v}\times\bm{B}) + q\Omega x \partial_y \bm{B} - q\Omega B_x \bm{e_y} + \eta\bm{\Delta{B}},\\
  \label{eq:base4}  \nabla \cdot \bm{v}&= 0,\\
  \label{eq:base5}  \nabla \cdot \bm{B}&= 0,
\end{align}
where $\bm{B}$ is the magnetic field, $\rho_0$ the density, $\nu$ the kinematic viscosity and $\eta$ the magnetic diffusivity. The gradient of the pressure perturbation $\bm{\nabla}\Pi$ is obtained from the constraint of a divergence-free
flow field (Eq.~\ref{eq:base4}). $\bm{v}$ is the velocity fluctuation with respect to the mean shear profile $\bm{U} = -q\Omega x \bm{e_y}$, where the shear parameter
$q \equiv - \dd \log \Omega / \dd \log r$ is assumed to have a sub-Keplerian value of $q=0.8$ in all simulations \citep[such value is relevant in particular for proto-neutron stars, e.g.][]{reboul-salze21,bugli20}. The box dimensions $(L_x,L_y,L_z) = (1,3,3) L$ are chosen in light of the comparison with global simulations performed by \citet{reboul-salze21}.

\begin{table*}
\caption{Overview of the numerical simulations. The second and third columns show the dimensionless control parameters that have been varied in this study: the Reynolds number $\mathrm{Re} \equiv L^2\Omega / \nu$ and the magnetic Prandtl number $\mathrm{Pm} \equiv \nu/\eta$. The time and volume averages of the kinetic ($E_{\rm kin} \equiv v^2/2$) and magnetic ($E_{\rm mag} \equiv B^2/2\mu_0$) energy densities and the total stress ($\alpha \equiv \rho_0 v_xv_y - B_xB_y/\mu_0$) are in units of $\rho_0 L^2\Omega^2$, with $L$ the radial size of the box with dimensions $(L_x,L_y,L_z) = (1,3,3) L$. The resistive length scale $l_{\rm res}$ is defined as the wavelength such that $50\%$ of the resistive dissipation takes place at shorter/longer scales. The ratio of the resistive scale to the radial size of the grid cells $\Delta x$ is used to check that the simulation is sufficiently resolved.}\label{table}
\begin{tabular}{lrrcccccrrrr}
\toprule
 Name & ${\rm Re}$ & ${\rm Pm}$ & $E_{\rm kin}$ & $E_{\rm mag}$ & $E_{\rm mag}^{k<3\pi/L}$ & $E_{\rm mag}^{k<\pi/L}$ & $\alpha$ & $n_x$ & $n_y$ & $n_z$ & $l_{\rm res}/\Delta x$ \\
\midrule
$ \text{Re400Pm24n64} $ & $400$ & $24$ & $ 2.89 \times 10^{-3} $ & $ 4.87 \times 10^{-2} $ & $ 9.14 \times 10^{-3} $ & $ 3.58 \times 10^{-3} $ & $ 1.90 \times 10^{-2} $ & $64$ & $96$ & $192$ & $ 8.4 $ \\
$ \text{Re400Pm24n96} $ & $400$ & $24$ & $ 3.85 \times 10^{-3} $ & $ 6.27 \times 10^{-2} $ & $ 1.42 \times 10^{-2} $ & $ 6.48 \times 10^{-3} $ & $ 2.45 \times 10^{-2} $ & $96$ & $128$ & $256$ & $ 12.1 $ \\
$ \text{Re400Pm32n96} $ & $400$ & $32$ & $ 4.14 \times 10^{-3} $ & $ 7.01 \times 10^{-2} $ & $ 1.38 \times 10^{-2} $ & $ 5.91 \times 10^{-3} $ & $ 2.75 \times 10^{-2} $ & $96$ & $128$ & $256$ & $ 10.4 $ \\
$ \text{Re400Pm32n96hyz} $ & $400$ & $32$ & $ 4.41 \times 10^{-3} $ & $ 7.51 \times 10^{-2} $ & $ 1.61 \times 10^{-2} $ & $ 7.27 \times 10^{-3} $ & $ 2.90 \times 10^{-2} $ & $96$ & $256$ & $512$ & $ 10.3 $ \\
$ \text{Re400Pm32n128} $ & $400$ & $32$ & $ 4.62 \times 10^{-3} $ & $ 7.70 \times 10^{-2} $ & $ 1.64 \times 10^{-2} $ & $ 7.77 \times 10^{-3} $ & $ 3.01 \times 10^{-2} $ & $128$ & $192$ & $384$ & $ 13.6 $ \\
$ \text{Re400Pm48n96} $ & $400$ & $48$ & $ 5.57 \times 10^{-3} $ & $ 1.03 \times 10^{-1} $ & $ 2.68 \times 10^{-2} $ & $ 1.79 \times 10^{-2} $ & $ 3.70 \times 10^{-2} $ & $96$ & $128$ & $256$ & $ 8.4 $ \\
$ \text{Re400Pm48n128} $ & $400$ & $48$ & $ 5.40 \times 10^{-3} $ & $ 9.77 \times 10^{-2} $ & $ 1.88 \times 10^{-2} $ & $ 9.57 \times 10^{-3} $ & $ 3.75 \times 10^{-2} $ & $128$ & $192$ & $384$ & $ 11.0 $ \\
$ \text{Re400Pm64n96} $ & $400$ & $64$ & $ 6.05 \times 10^{-3} $ & $ 1.10 \times 10^{-1} $ & $ 2.46 \times 10^{-2} $ & $ 1.30 \times 10^{-2} $ & $ 4.03 \times 10^{-2} $ & $96$ & $128$ & $256$ & $ 7.5 $ \\
$ \text{Re400Pm64n128} $ & $400$ & $64$ & $ 5.98 \times 10^{-3} $ & $ 1.17 \times 10^{-1} $ & $ 2.54 \times 10^{-2} $ & $ 1.51 \times 10^{-2} $ & $ 4.28 \times 10^{-2} $ & $128$ & $192$ & $384$ & $ 9.5 $ \\
$ \text{Re400Pm80n128} $ & $400$ & $80$ & $ 6.55 \times 10^{-3} $ & $ 1.33 \times 10^{-1} $ & $ 2.87 \times 10^{-2} $ & $ 1.69 \times 10^{-2} $ & $ 4.75 \times 10^{-2} $ & $128$ & $192$ & $384$ & $ 8.5 $ \\
$ \text{Re400Pm80n192} $ & $400$ & $80$ & $ 6.24 \times 10^{-3} $ & $ 1.35 \times 10^{-1} $ & $ 3.00 \times 10^{-2} $ & $ 1.70 \times 10^{-2} $ & $ 4.69 \times 10^{-2} $ & $192$ & $256$ & $512$ & $ 12.6 $ \\
$ \text{Re400Pm96n128} $ & $400$ & $96$ & $ 6.75 \times 10^{-3} $ & $ 1.41 \times 10^{-1} $ & $ 3.15 \times 10^{-2} $ & $ 1.76 \times 10^{-2} $ & $ 4.93 \times 10^{-2} $ & $128$ & $192$ & $384$ & $ 7.9 $ \\
$ \text{Re400Pm96n192} $ & $400$ & $96$ & $ 6.69 \times 10^{-3} $ & $ 1.45 \times 10^{-1} $ & $ 2.58 \times 10^{-2} $ & $ 1.06 \times 10^{-2} $ & $ 5.23 \times 10^{-2} $ & $192$ & $256$ & $512$ & $ 11.5 $ \\
$ \text{Re400Pm128n128} $ & $400$ & $128$ & $ 6.76 \times 10^{-3} $ & $ 1.58 \times 10^{-1} $ & $ 4.42 \times 10^{-2} $ & $ 2.94 \times 10^{-2} $ & $ 4.98 \times 10^{-2} $ & $128$ & $192$ & $384$ & $ 7.1 $ \\
$ \text{Re400Pm128n192} $ & $400$ & $128$ & $ 7.04 \times 10^{-3} $ & $ 1.61 \times 10^{-1} $ & $ 3.28 \times 10^{-2} $ & $ 2.14 \times 10^{-2} $ & $ 5.56 \times 10^{-2} $ & $192$ & $256$ & $512$ & $ 10.0 $ \\
$ \text{Re400Pm192n256} $ & $400$ & $192$ & $ 7.56 \times 10^{-3} $ & $ 2.03 \times 10^{-1} $ & $ 3.69 \times 10^{-2} $ & $ 1.80 \times 10^{-2} $ & $ 6.69 \times 10^{-2} $ & $256$ & $384$ & $768$ & $ 10.8 $ \\
$ \text{Re400Pm256n256} $ & $400$ & $256$ & $ 7.23 \times 10^{-3} $ & $ 2.13 \times 10^{-1} $ & $ 4.93 \times 10^{-2} $ & $ 3.59 \times 10^{-2} $ & $ 6.57 \times 10^{-2} $ & $256$ & $384$ & $768$ & $ 9.6 $ \\
$ \text{Re800Pm10n96} $ & $800$ & $10$ & $ 2.31 \times 10^{-3} $ & $ 2.87 \times 10^{-2} $ & $ 5.99 \times 10^{-3} $ & $ 2.74 \times 10^{-3} $ & $ 1.17 \times 10^{-2} $ & $96$ & $128$ & $256$ & $ 13.0 $ \\
$ \text{Re800Pm12.5n96} $ & $800$ & $12$ & $ 2.43 \times 10^{-3} $ & $ 3.10 \times 10^{-2} $ & $ 5.26 \times 10^{-3} $ & $ 2.28 \times 10^{-3} $ & $ 1.30 \times 10^{-2} $ & $96$ & $128$ & $256$ & $ 11.5 $ \\
$ \text{Re800Pm16n128} $ & $800$ & $16$ & $ 3.46 \times 10^{-3} $ & $ 4.35 \times 10^{-2} $ & $ 7.66 \times 10^{-3} $ & $ 3.52 \times 10^{-3} $ & $ 1.86 \times 10^{-2} $ & $128$ & $192$ & $384$ & $ 13.0 $ \\
$ \text{Re800Pm22.5n128} $ & $800$ & $22$ & $ 4.79 \times 10^{-3} $ & $ 6.06 \times 10^{-2} $ & $ 1.05 \times 10^{-2} $ & $ 4.64 \times 10^{-3} $ & $ 2.62 \times 10^{-2} $ & $128$ & $192$ & $384$ & $ 10.5 $ \\
$ \text{Re800Pm32n128} $ & $800$ & $32$ & $ 6.26 \times 10^{-3} $ & $ 8.03 \times 10^{-2} $ & $ 1.47 \times 10^{-2} $ & $ 7.45 \times 10^{-3} $ & $ 3.42 \times 10^{-2} $ & $128$ & $192$ & $384$ & $ 8.7 $ \\
$ \text{Re800Pm48n192} $ & $800$ & $48$ & $ 7.18 \times 10^{-3} $ & $ 1.01 \times 10^{-1} $ & $ 1.84 \times 10^{-2} $ & $ 8.27 \times 10^{-3} $ & $ 4.20 \times 10^{-2} $ & $192$ & $256$ & $512$ & $ 10.4 $ \\
$ \text{Re800Pm64n256} $ & $800$ & $64$ & $ 8.17 \times 10^{-3} $ & $ 1.22 \times 10^{-1} $ & $ 2.26 \times 10^{-2} $ & $ 1.35 \times 10^{-2} $ & $ 4.98 \times 10^{-2} $ & $256$ & $384$ & $768$ & $ 11.8 $ \\
$ \text{Re800Pm80n256} $ & $800$ & $80$ & $ 8.10 \times 10^{-3} $ & $ 1.25 \times 10^{-1} $ & $ 1.35 \times 10^{-2} $ & $ 3.82 \times 10^{-3} $ & $ 5.12 \times 10^{-2} $ & $256$ & $384$ & $768$ & $ 10.5 $ \\
$ \text{Re800Pm96n256} $ & $800$ & $96$ & $ 8.69 \times 10^{-3} $ & $ 1.47 \times 10^{-1} $ & $ 2.83 \times 10^{-2} $ & $ 1.51 \times 10^{-2} $ & $ 5.52 \times 10^{-2} $ & $256$ & $384$ & $768$ & $ 9.7 $ \\
$ \text{Re1600Pm8n96} $ & $1600$ & $8$ & $ 1.03 \times 10^{-3} $ & $ 1.25 \times 10^{-2} $ & $ 1.71 \times 10^{-3} $ & $ 7.30 \times 10^{-4} $ & $ 4.93 \times 10^{-3} $ & $96$ & $128$ & $256$ & $ 10.4 $ \\
$ \text{Re1600Pm10n128} $ & $1600$ & $10$ & $ 1.37 \times 10^{-3} $ & $ 1.60 \times 10^{-2} $ & $ 1.89 \times 10^{-3} $ & $ 7.83 \times 10^{-4} $ & $ 6.74 \times 10^{-3} $ & $128$ & $192$ & $384$ & $ 11.8 $ \\
$ \text{Re1600Pm12.5n128} $ & $1600$ & $12$ & $ 3.26 \times 10^{-3} $ & $ 3.34 \times 10^{-2} $ & $ 5.39 \times 10^{-3} $ & $ 2.76 \times 10^{-3} $ & $ 1.49 \times 10^{-2} $ & $128$ & $192$ & $384$ & $ 9.7 $ \\
$ \text{Re1600Pm16n128} $ & $1600$ & $16$ & $ 3.98 \times 10^{-3} $ & $ 4.06 \times 10^{-2} $ & $ 6.36 \times 10^{-3} $ & $ 3.17 \times 10^{-3} $ & $ 1.83 \times 10^{-2} $ & $128$ & $192$ & $384$ & $ 8.5 $ \\
$ \text{Re1600Pm24n192} $ & $1600$ & $24$ & $ 6.14 \times 10^{-3} $ & $ 6.39 \times 10^{-2} $ & $ 1.04 \times 10^{-2} $ & $ 4.93 \times 10^{-3} $ & $ 2.87 \times 10^{-2} $ & $192$ & $256$ & $512$ & $ 9.8 $ \\
$ \text{Re1600Pm32n256} $ & $1600$ & $32$ & $ 5.40 \times 10^{-3} $ & $ 6.27 \times 10^{-2} $ & $ 7.47 \times 10^{-3} $ & $ 2.68 \times 10^{-3} $ & $ 2.76 \times 10^{-2} $ & $256$ & $384$ & $768$ & $ 11.2 $ \\
$ \text{Re1600Pm48n256} $ & $1600$ & $48$ & $ 8.84 \times 10^{-3} $ & $ 1.10 \times 10^{-1} $ & $ 2.50 \times 10^{-2} $ & $ 1.51 \times 10^{-2} $ & $ 4.32 \times 10^{-2} $ & $256$ & $384$ & $768$ & $ 8.9 $ \\
\bottomrule
\end{tabular}
\end{table*}

\begin{figure}
    \centering
    \includegraphics[width=\columnwidth]{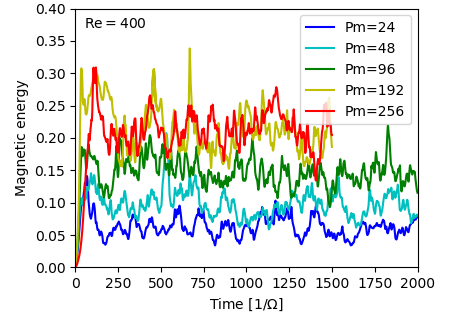}
   \includegraphics[width=\columnwidth]{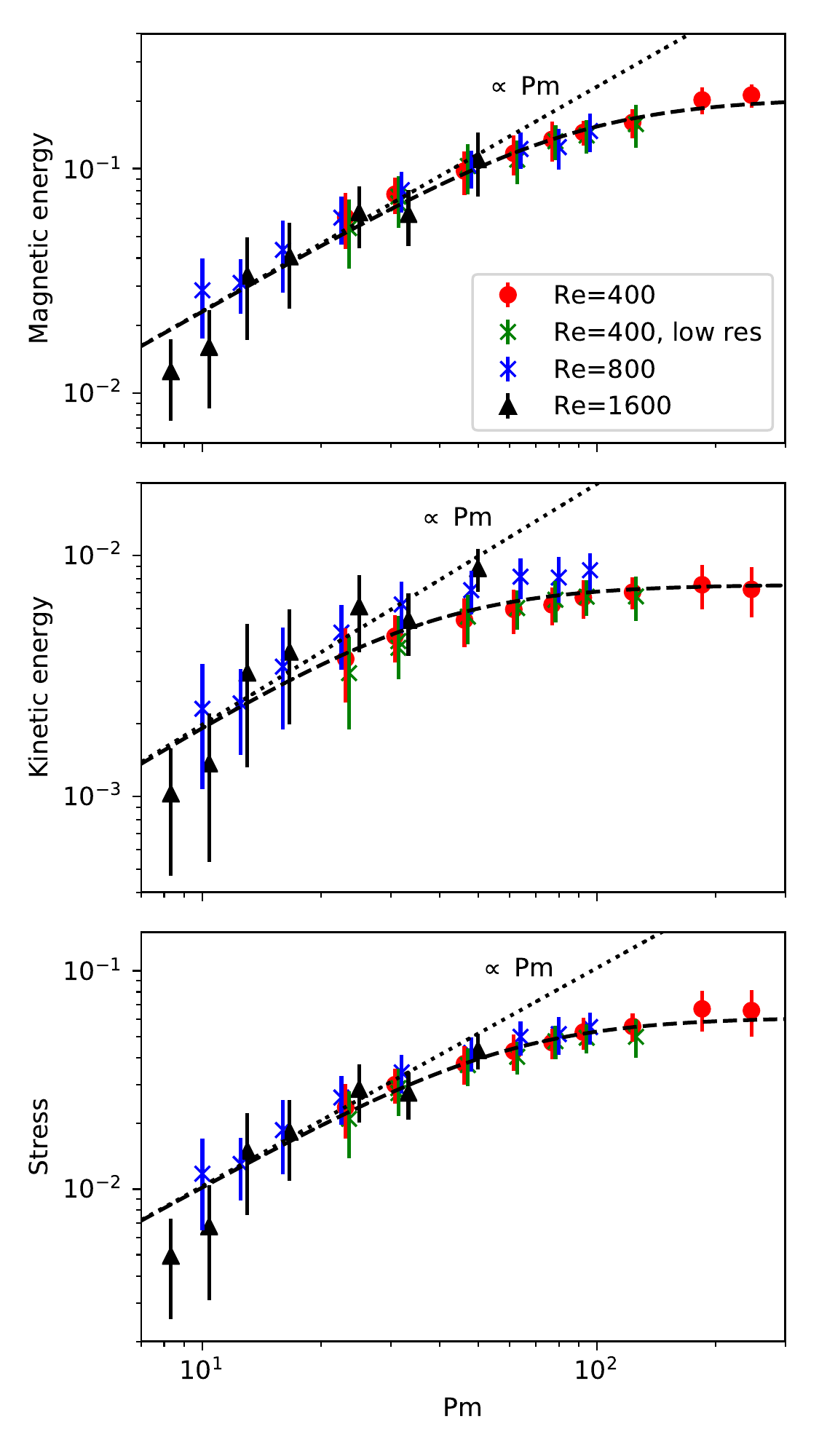}   \caption{Top panel: Time evolution of the magnetic energy for a subset of simulations with $\mathrm{Re}=400$ and $\mathrm{Pm}\in[24,256]$. Lower panels: Time and volume averaged magnetic energy ($E_{\rm mag} \equiv B^2/2\mu_0$), kinetic energy ($E_{\rm kin}\equiv v^2/2$) and stress ($\alpha \equiv \rho_0 v_xv_y - B_xB_y/\mu_0$) as functions of the magnetic Prandtl number for three different values of the Reynolds number $\mathrm{Re}=400$ (red circles), $\mathrm{Re}=800$ (blue crosses) and $\mathrm{Re}=1600$ (black triangles). The error bars show the standard deviation, which probably overestimates the actual error on the averages. The dotted line shows a linear fit valid at moderate $\mathrm{Pm}$, while the dashed line shows the fit with \refeq{eq:fit}. The stress and energies are in units of $\rho_0 L^2\Omega^2$ (with $L$ the radial size of the box).}
    \label{fig:energy}
\end{figure}

\begin{figure*}
    \centering
    \includegraphics[trim={0 1.35cm 9.1cm 3.1cm},clip,width=0.28\textwidth]{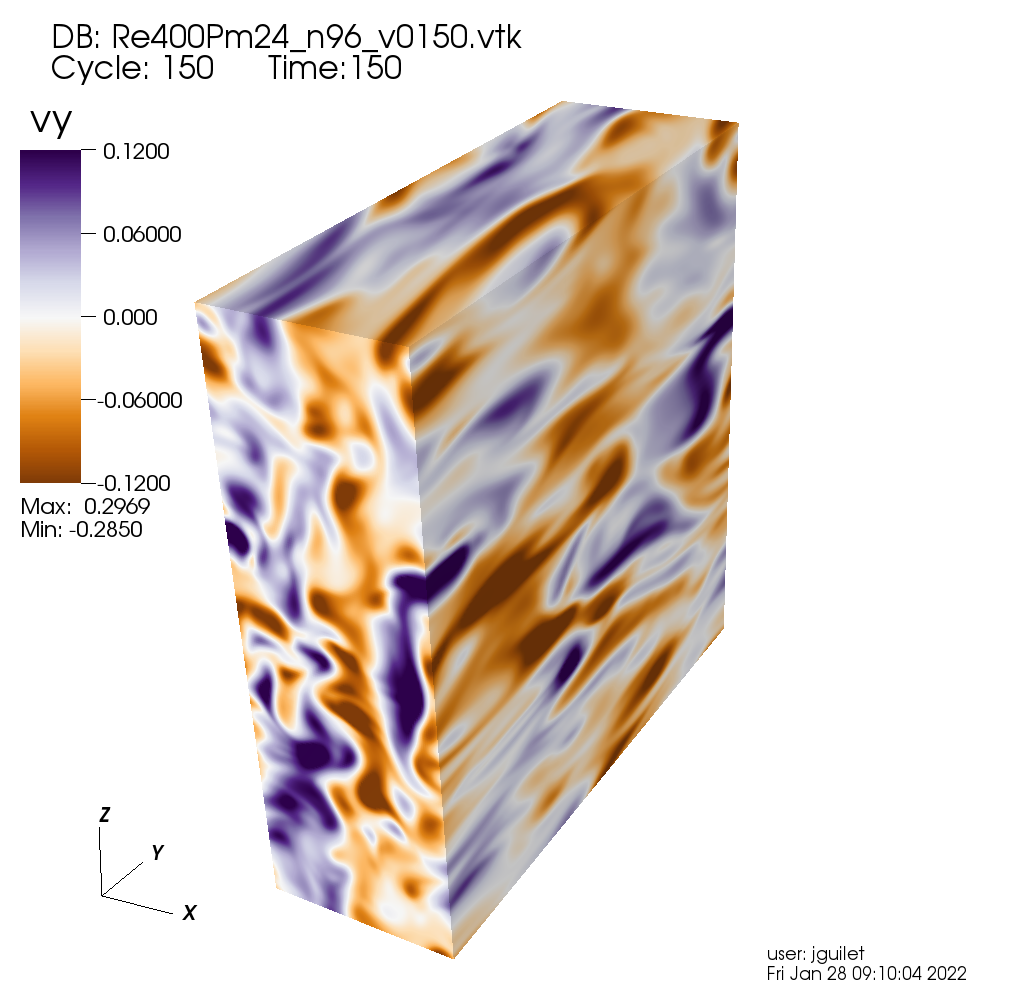} 
    \includegraphics[trim={0 1.35cm 9.1cm 3.1cm},clip,width=0.28\textwidth]{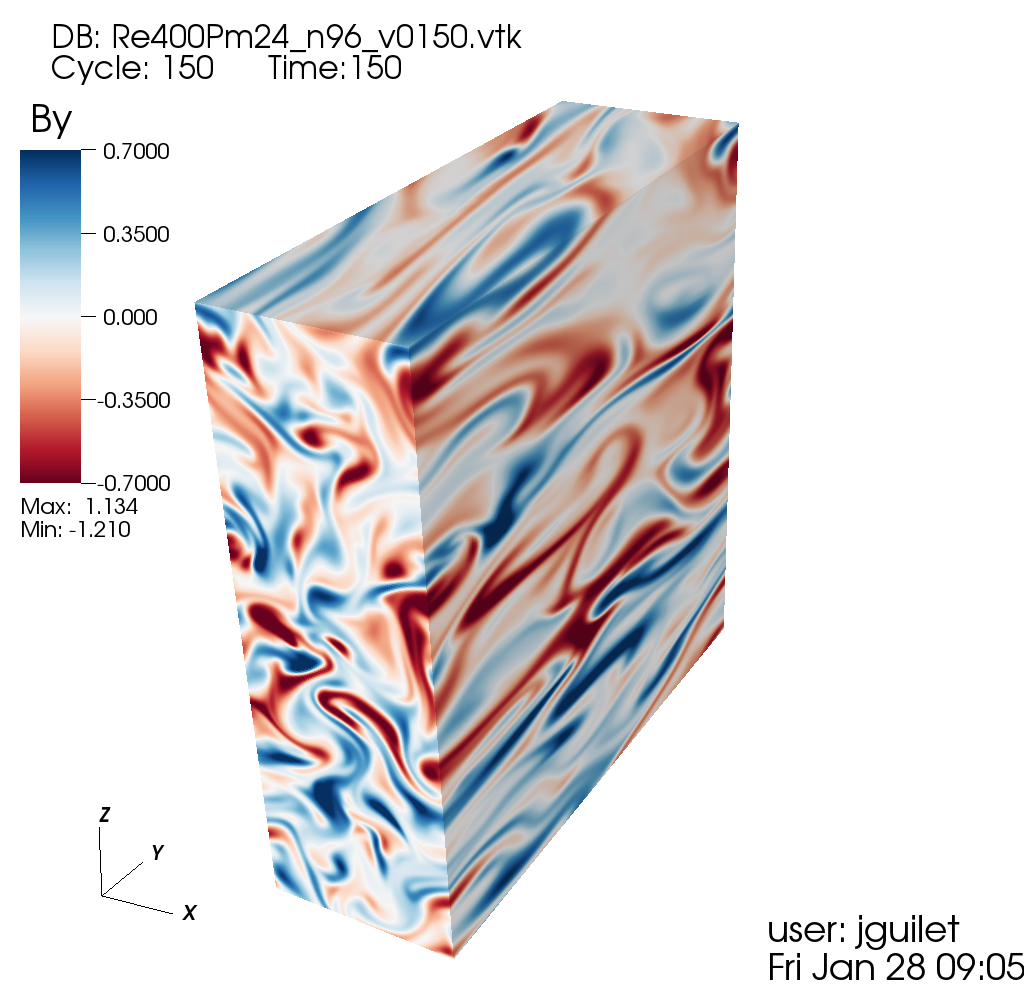}
    \includegraphics[width=0.43\textwidth]{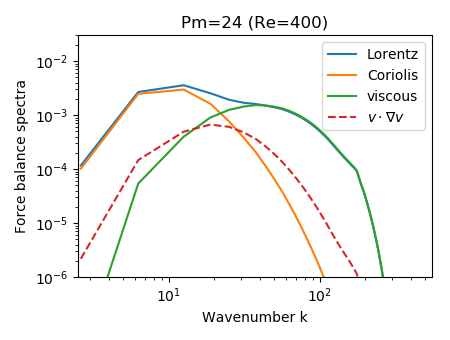}
    \includegraphics[trim={0 2.7cm 18.2cm 6.2cm},clip,width=0.28\textwidth]{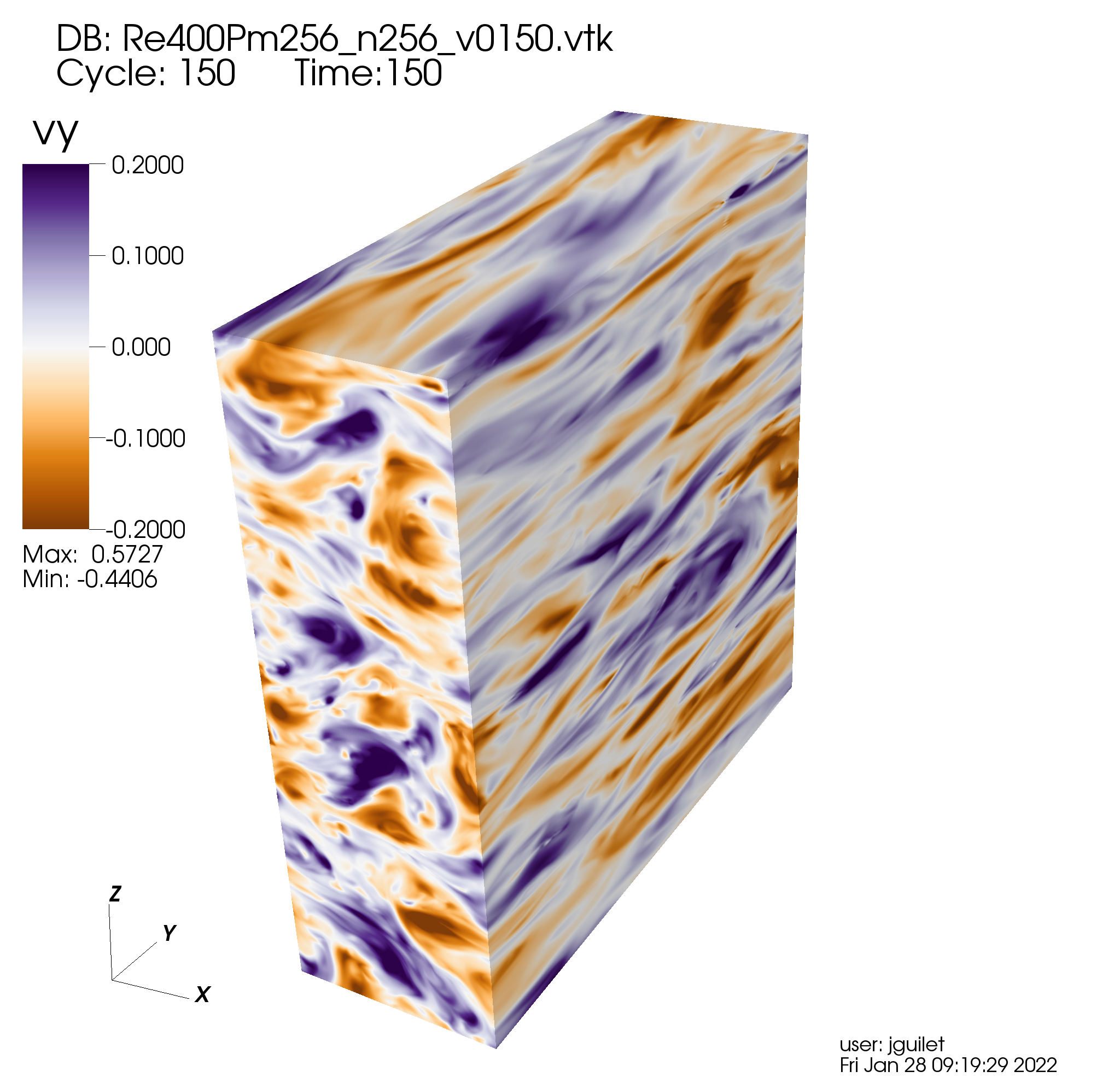}
    \includegraphics[trim={0 2.7cm 18.2cm 6.2cm},clip,width=0.28\textwidth]{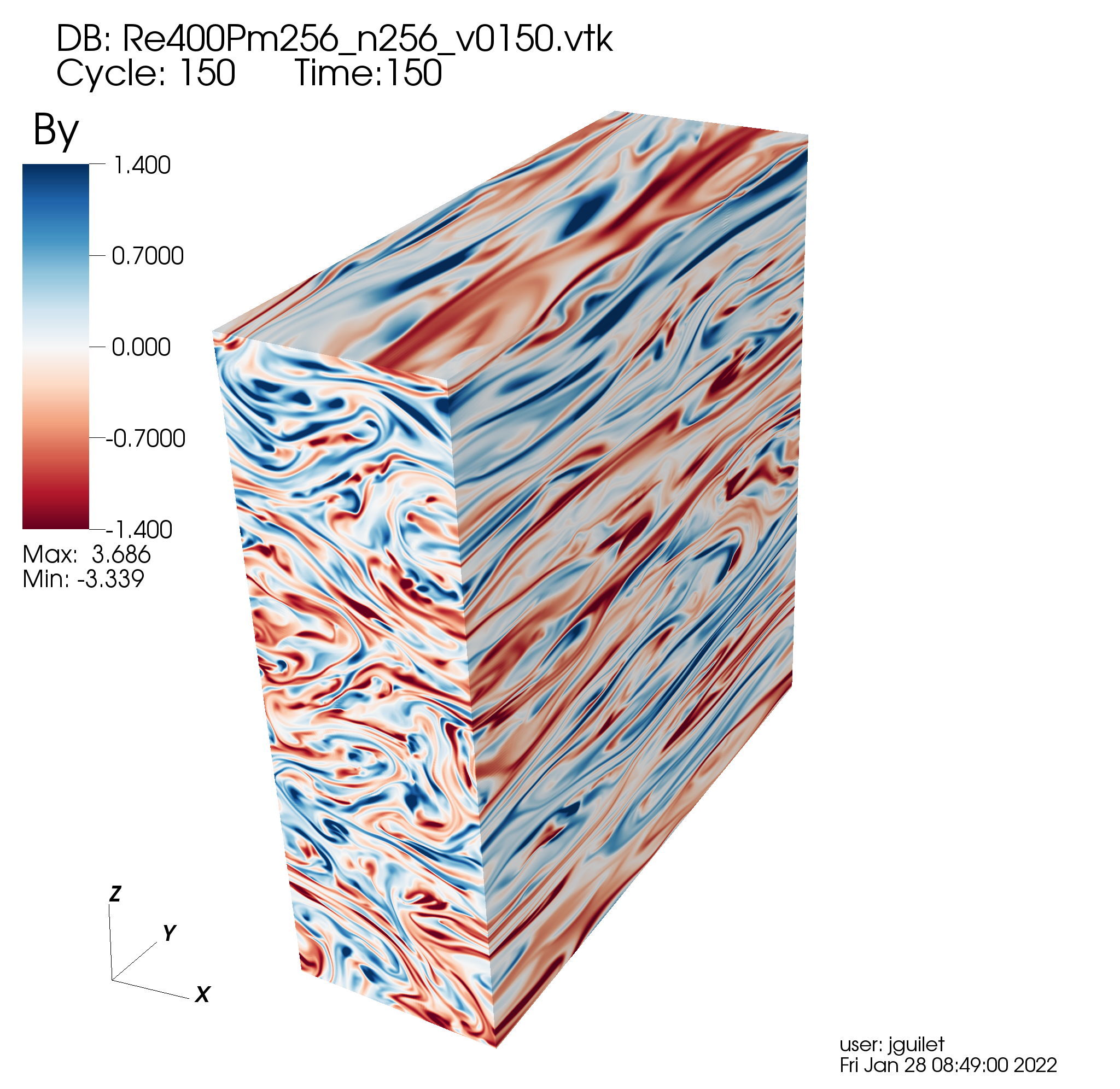}
    \includegraphics[width=0.43\textwidth]{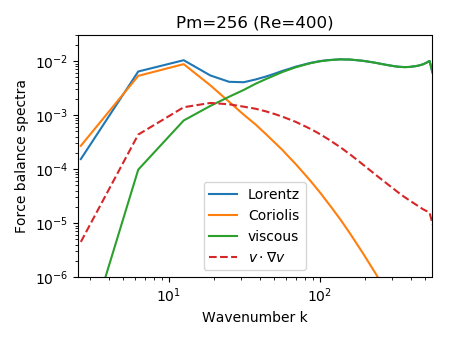}
    \caption{3D rendering of the azimuthal component of the velocity field (left panels) and of the magnetic field (middle panels). The right panels represent the time-averaged spectra of the force balance. The upper row corresponds to a simulation with a moderate value of $\mathrm{Pm}=24$, while the lower row corresponds to a large value of $\mathrm{Pm}=256$. Both simulations share the same Reynolds number $\mathrm{Re}=400$. In the force balance spectrum, only the divergence-free part of the forces is included as it is appropriate in the incompressible approximation.}
    \label{fig:3d}
\end{figure*}

The shear rate and the box aspect ratio being fixed, this setup is governed by only two dimensionless numbers: the {\it Reynolds number} $\mathrm{Re} \equiv L^2\Omega / \nu$ and the {\it magnetic Reynolds number} $\mathrm{Rm} \equiv L^2\Omega / \eta$. We obtained self-sustained MRI-driven turbulence in 30 simulations with $\mathrm{Rm}$ ranging from $8\, 000$ to $102\, 400$ and with three different values of the Reynolds number ($\mathrm{Re}=400$, $800$ and $1\,600$). In this set of simulations, the magnetic Prandtl number $\mathrm{Pm} \equiv \nu/\eta = \mathrm{Rm/Re} $ lies in the range $\mathrm{Pm} = 8-256$. Note that for each Reynolds number, we ran simulations with lower magnetic Reynolds number where the dynamo was not self-sustained, which are therefore not included in the present analysis.

Our simulations are initialised with a random superposition of large-scale magnetic modes, with exactly zero net magnetic flux. We checked in a few cases that, after a transient phase, the turbulent state was statistically independent of the initial conditions, provided that the initial magnetic field was strong enough to initiate an MRI-driven dynamo. The simulations were run for a duration varying between $1500$ and $3000\, \Omega^{-1}$, {which is longer than the typical timescale of the fluctuations} (see upper panel of \refig{fig:energy}). The time-averaged results shown in this paper are performed at times $t>500 \, \Omega^{-1}$ in order to exclude any initial transient behaviour. 

Throughout the paper, our results are normalised using the radial size
of the domain $L$, the angular frequency
$\Omega$, and the density
$\rho_0$. With parameters typical of a proto-neutron star $L=10\,{\rm km}$,
$\Omega = 10^3\,{\rm s^{-1}}$, and
$\rho_0 = 2\times 10^{13}\,{\rm g\,cm^{-3}}$, time would therefore be
measured in units of $1\,{\rm ms}$, velocity in units of
$10^9 \,{\rm cm\,s^{-1}}$, the magnetic field in units of
$1.6\times10^{16}\,{\rm G}$, and  the energy density in units of
$2\times 10^{31}\,{\rm erg\,cm^{-3}}$. Our range of Reynolds numbers translates into a viscosity ranging from $\nu = 6\times 10^{11}$ to  $2.5\times 10^{12} \,{\rm cm^2\,s^{-1}}$, which is comparable to estimates of the neutrino viscosity in the outer parts of a PNS \citep{guilet15}.

\subsection{Numerical methods and convergence tests}
In order to solve the incompressible MHD
equations~(\ref{eq:base1})--(\ref{eq:base5}), we use the
pseudo-spectral code \textsc{snoopy} \citep{lesur05,lesur07}, which has been used in numerous studies of the MRI \citep[e.g.][]{lesur11,guilet15,walker17}. 
Our simulations were performed using a grid resolution varying from $(n_x,n_y,n_z)=(64,128,192)$ to $(n_x,n_y,n_z)=(256,512,768)$. In order to ensure that the resistive scale is resolved, the number of grid points is increased as Pm is increased. In all simulations presented in the figures of this paper {(except for some of the low resolution simulations performed for the convergence study)}, the resistive scale is resolved by at least 8.5 grid cells (see Table~\ref{table}). 

{In order to check the convergence of our results, we ran a subset of {8} additional simulations at lower resolution for $\mathrm{Re}=400$, $\mathrm{Pm}=24-128$. We define a convergence criterion $l_{\rm res}/\Delta x$ as the ratio of the resistive scale (measured such that half of the resistive dissipation takes place at higher/lower scales) to our grid scale. The precise threshold needed for convergence is expected to depend on the numerical scheme\footnote{Pseudo-spectral methods like the one we use are known for their low dissipation and for necessitating fewer grid points than non-spectral grid-based methods \citep[e.g.][]{fromang07b}} and potentially on the physical problem considered through the width of the dissipation peak. As a consequence, it is necessary to calibrate it through dedicated convergence tests as it is done here. In all our simulations, the wavenumber at which the dissipation peak has decreased by a factor 2 is roughly twice as large as the wavenumber of the dissipation maximum independently of Pm. This suggests that a fixed threshold in $l_{\rm res}/\Delta x$ can be a meaningful measure of convergence in the parameter space explored. When comparing the low resolution runs with their higher resolution counterparts (Table~\ref{table} and Fig.~\ref{fig:energy}), we observe that the stress and turbulent energies of the low resolution simulations have a tendency to be slightly smaller (with the exception of a few cases) but they are all consistent within the statistical error bars defined with the standard deviation. Most simulations show differences of at most a few percent, while the least resolved test (Re400Pm128n128 with $l_{\rm res}/\Delta x = 7.1$, which we would deem slightly under-resolved) has a larger difference of about $10\%$ in the stress compared to its high resolution counterpart (however, the difference is smaller for the turbulent energies). We stress that all our runs are better resolved than this under-resolved test according to our resolution criterion ($l_{\rm res}/\Delta x > 8.5$ for all simulations, and the three highest Pm runs have $l_{\rm res}/\Delta x > 9.6$). Our convergence tests therefore suggest that all the simulations included in the analysis give reliable results on the average energies and stress with errors of at most a few percent due to numerical artefacts. Such errors are within the statistical errors shown in Fig.~\ref{fig:energy} and do not compromise our main conclusions.} 

{In some of the spectra, a numerical artefact can be discerned at wavenumbers larger than the dissipation peak. We checked that, in all the simulations included in the analysis, the resistive dissipation rate at this artefact is smaller than the  resistive dissipation peak by a factor of at least 10 (e.g. 40 and 15 for the simulations at $\mathrm{Pm}=192$ and 256, respectively). In four of the low resolution tests this artefact is more pronounced and leads to a breakdown of this criterion (Re400Pm48n96, Re400Pm64n96, Re400Pm96n128, Re400Pm128n128 with a ratio of the artefact to peak resistive dissipation rate of 6, 3, 7 and 4, respectively), which allows to check the potential influence of such artefact on the overall dynamics and energetics. Therefore, the convergence tests discussed above give confidence that the influence of the sub-resistive-scale artefact on the averaged energies and stress is small (at most a few percent) in the simulations included in the analysis.}  

Finally, note that the resolution in the azimuthal direction is twice as low as in the radial or vertical direction, because the structures are more elongated in the azimuthal direction due to the shear (\refig{fig:3d}). We checked in a few cases that our results are not affected by the lower azimuthal resolution. 

\section{Results}
	\label{sec:results}

\begin{figure}
    \centering
     \includegraphics[width=\columnwidth]{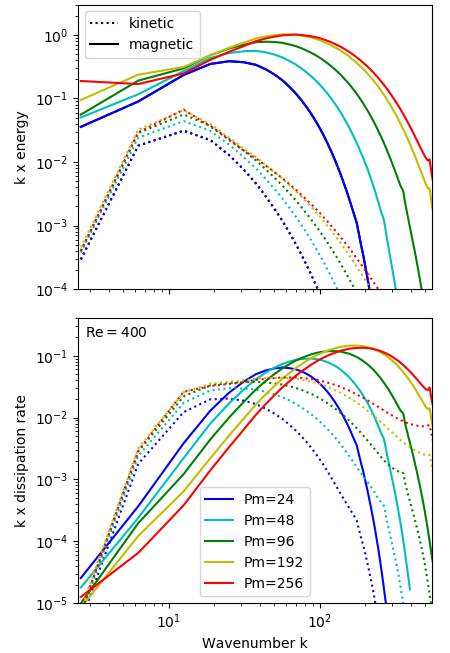}
    \caption{Time-averaged spectra of a subset of simulations with $\mathrm{Re}=400$ and varying $\mathrm{Pm}$. Upper panel: kinetic (dotted lines) and magnetic (solid lines) energy spectra. Lower panel: viscous (dotted lines) and resistive (solid lines) dissipation rate spectra. All spectra are multiplied by $k$ so that the typical scale containing most of the energy or dissipation is clearly visualised as the maximum of the curve.}
    \label{fig:spectra}
\end{figure}

\begin{figure}
    \centering
     \includegraphics[width=\columnwidth]{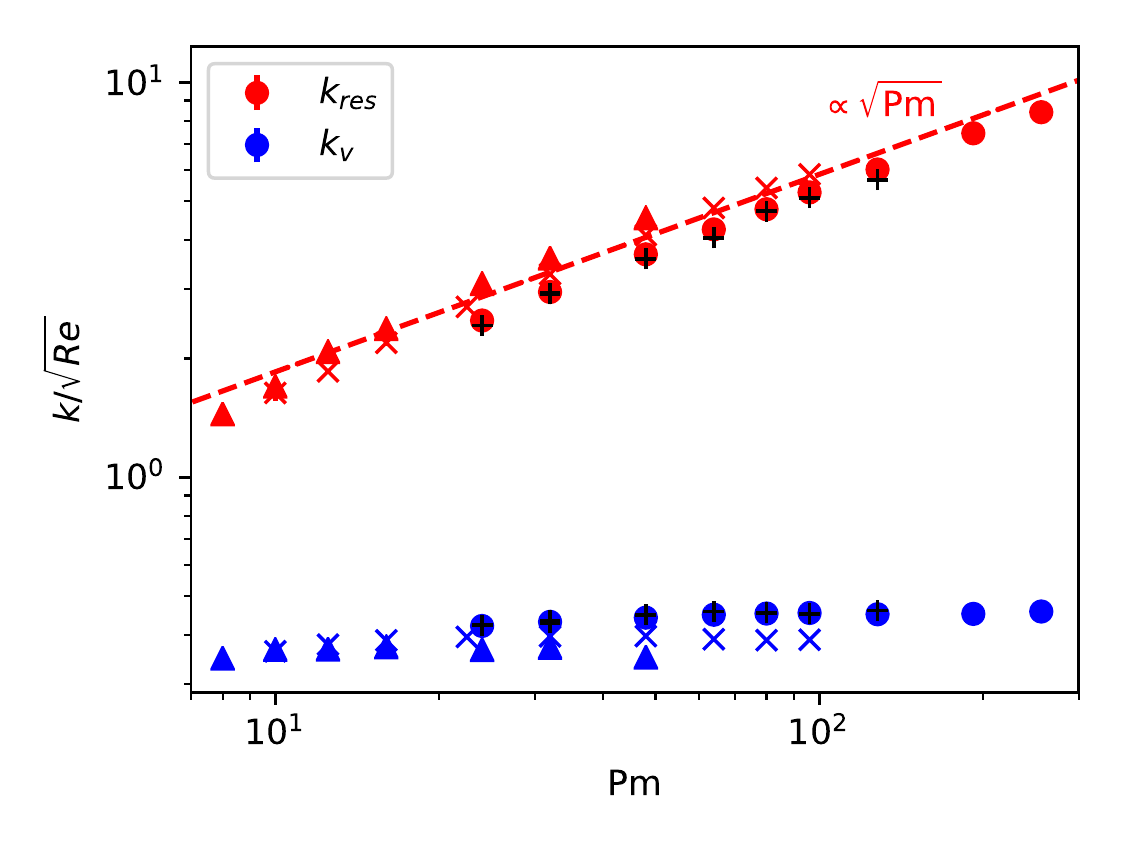}
    \caption{Typical wavenumbers of the resistive dissipation (red symbols, defined such that half of the resistive dissipation takes place at higher/lower wavenumber) and of the kinetic energy (blue, defined such that half of the kinetic energy is located at higher/lower wavenumbers). The wavenumbers have been normalised by $\sqrt{\mathrm{Re}}$ so that the dependence on the Reynolds number is approximately scaled out. The symbol shapes correspond to different Reynolds numbers as in \refig{fig:energy}: $\mathrm{Re}=400$ (circles), $\mathrm{Re}=800$ (crosses) and $\mathrm{Re}=1600$ (triangles). The low resolution tests at $\mathrm{Re}=400$ are represented with the black plus signs. The dashed line represent a scaling of the resistive wavenumber as $\propto \sqrt{\mathrm{Pm}}$.}
    \label{fig:wavenumbers}
\end{figure}

\begin{figure}
    \centering
     \includegraphics[width=\columnwidth]{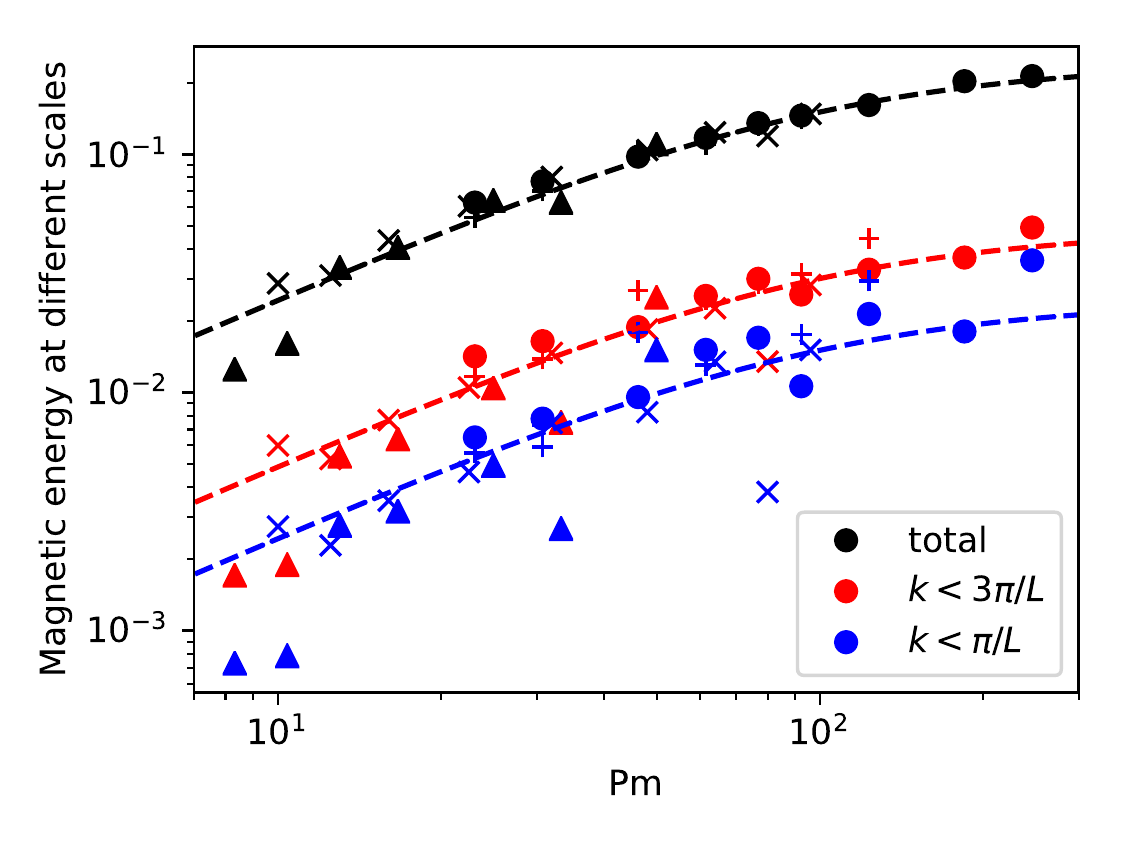}
    \caption{Magnetic energy contained in the largest scales of the box as a function of the magnetic Prandtl number. Blue and red symbols correspond to the energy at wavenumbers smaller than $\pi/L$ and $3\pi/L$ respectively, while black symbols show the total magnetic energy. The symbol shapes correspond to different Reynolds numbers as in \refig{fig:energy}: $\mathrm{Re}=400$ (circles), $\mathrm{Re}=800$ (crosses) and $\mathrm{Re}=1600$ (triangles). The low resolution tests at $\mathrm{Re}=400$ are represented with the plus signs. The dashed lines represent the fit of the total magnetic energy (black), and a fraction of $20\%$ (red) and $10\%$ (blue) of this fit, respectively.}
    \label{fig:large-scale_B}
\end{figure}

The time-evolution of the magnetic energy for a subset of our simulations with varying $\mathrm{Pm}$ shows a clear trend of increasing magnetic energy for larger Pm (\refig{fig:energy}, upper panel). This trend nevertheless stops towards the largest values of $\mathrm{Pm}$, as the two simulations $\mathrm{Pm}=192$ and $\mathrm{Pm}=256$ appear to converge on the same value. This conclusion is confirmed quantitatively by the lower panels of \refig{fig:energy}, showing the magnetic energy, kinetic energy and total stress time-averaged over the quasi-stationary turbulent phase. All three quantities increase approximately linearly with Pm up to $\mathrm{Pm} \lesssim 50$. Interestingly, at still larger values of $\mathrm{Pm}$, they transition to a less steep dependence and are consistent with a plateau for $\mathrm{Pm} \gtrsim 100$ {(or slightly more for the magnetic energy)}. It is noteworthy that all three sets of simulations with different Reynolds numbers fall on the same universal curves as a function of $\mathrm{Pm}$. This dependence can be well fitted by a functional form 
\begin{equation}
A(\mathrm{Pm}) = A^\infty \frac{\mathrm{Pm}/\mathrm{Pm_c}}{\sqrt{1+(\mathrm{Pm}/\mathrm{Pm_c})^2}},
    \label{eq:fit}
\end{equation}
as represented by dotted lines in \refig{fig:energy}. The best fit parameters give critical magnetic Prandtl numbers for the start of the plateau of $\mathrm{Pm_c}=91$, $37$ and $61$ for the magnetic energy, kinetic energy and the stress respectively. The asymptotic values at large $\mathrm{Pm}$ are $E^{\infty}_\mathrm{mag} = 0.21$, $E^{\infty}_\mathrm{kin}=7.8\times 10^{-3}$ and $\alpha^\infty = 6.4\times 10^{-2}$. Because of the higher value of $\mathrm{Pm_c}$ for the magnetic energy, Pm-independent plateaus are more clearly visible for the stress and kinetic energy. As a consequence, we cannot rule out a slight increase of the magnetic energy with Pm at high Pm, either logarithmically or in a very weak power-law \citep[similarly to][]{alexakis11}. 

A striking feature of Fig.~\ref{fig:energy} is that, for a given Pm, the stress and magnetic energy are independent of the Reynolds number and that the kinetic energy has only a weak dependence with Re. This may seem surprising given the relatively low values of Re considered, which could have suggested that the viscosity would impact the MRI. The impact of viscosity on the MRI linear growth is controlled by the viscous Elsasser number $E_\nu \equiv v_A^2/\nu\Omega$, with strong viscous effects for $E_\nu <1$. Although there is no linear phase of the MRI for the zero-net flux case, we can try to estimate the impact of viscosity on the saturated state with the Elsasser number computed with the turbulent magnetic field. In our simulations, $E_\nu$ is found to range from 40 to 350. Such relatively high values may give a hint as to why the stress and energies are roughly independent of the Reynolds number.

In MRI-driven turbulence, the turbulent magnetic energy is usually significantly larger than the kinetic energy, and similarly the Maxwell stress is larger than the Reynolds stress. In a Keplerian disk, the magnetic to kinetic ratios are often in the range 3 to 6, though they may depend on the precise setup. In our simulations, the kinetic to magnetic ratios are higher (between 10 and 20), which is likely due to the sub-Keplerian shear \citep[see Figure 4 of][]{pessah06a}.

The structures of the velocity and magnetic field are illustrated in \refig{fig:3d} with snapshots representative of the regimes of moderate $\mathrm{Pm}$ (upper row) and large $\mathrm{Pm}$ (lower row). The first striking feature is the highly entangled small-scale structure of the magnetic field in the high $\mathrm{Pm}$ regime. By contrast, in both the moderate and high $\mathrm{Pm}$ regimes the velocity field is dominated by structures at much larger scales comparable to the radial size of the box. One can nonetheless note the appearance of subdominant small-scales structures in the velocity field in the high $\mathrm{Pm}$ regime. 

A more quantitative description of the velocity and magnetic field structures is provided by their energy and dissipation spectra shown in \refig{fig:spectra} for varying values of $\mathrm{Pm}$ at $\mathrm{Re}=400$. As $\mathrm{Pm}$ increases, both the magnetic and kinetic energy spectra initially increase \textit{at all wavenumbers}\footnote{The only exception being the intermediate range of wavenumbers for $\mathrm{Pm}=256$.}. 
The kinetic energy peaks at a small wavenumber $k\sim 10$ independently of Pm, while the peak of the magnetic energy spectrum increases with Pm from $k\sim30$ to $k\sim 70$. Although the peak of the kinetic energy does not change with Pm, a significant tail at higher wavenumbers appears in the high Pm regime. This can be understood as a result of a balance between the Lorentz and viscous forces, which drives fluid motions (\refig{fig:3d}, right panels). The viscous dissipation in this tail broadens significantly the peak of viscous dissipation across larger wavenumbers (lower panel of \refig{fig:spectra}). Nevertheless, the majority of the dissipation is due to the resistivity acting at still higher wavenumbers, with a peak of the dissipation rate ranging from $k\sim 60$ to $k\sim 200$ as $\mathrm{Pm}$ increases. 

{Fig.~\ref{fig:wavenumbers} shows the Pm dependence of the characteristic scales of resistive dissipation (red symbols) and kinetic energy (blue symbols), where the dependence on the Reynolds number has been approximately scaled out by dividing the wavenumbers by $\sqrt{\mathrm{Re}}$. The characteristic resistive wavenumber is proportional to $\sqrt{\mathrm{Pm}}$ while the kinetic energy scale is approximately independent of Pm. The scaling of the resistive wavenumber may be explained by equating the shearing term that generates magnetic field at a rate $q \Omega$ and the resistive decay rate $\eta k^2$, which leads to an estimate of the resistive wavenumber $kL \sim \sqrt{\mathrm{Rm}}$ or equivalently $kL/\sqrt{\mathrm{Re}} \sim \sqrt{\mathrm{Pm}}$.} 

The analysis of the spectra {and characteristic scales} may suggest the following interpretation for the transition to a plateau independent of Pm. As Pm is increased (with fixed Re), the resistive scale becomes shorter and shorter as $\propto \sqrt{\mathrm{Pm}}$ while the velocity scale stays constant (see Fig.~\ref{fig:spectra} and Fig.~\ref{fig:wavenumbers}). At very high $\mathrm{Pm} >100$, the resistive scale is at least an order of magnitude shorter than the velocity scale. The dynamics may then become independent of the resistivity because the scale separation between the velocity and resistive scales prevents an efficient feedback of the resistive scales on the much larger scales of the flow. This situation would then be analogous to hydrodynamic turbulence, whose large-scale dynamics and overall energy budget become independent of the Reynolds number when a sufficient scale separation prevents feedback of the viscous scales on the injection scales.

We stress again that increasing $\mathrm{Pm}$ not only leads to more intense smaller-scale magnetic fields but also to stronger large-scale magnetic fields. \refig{fig:large-scale_B} shows that the magnetic energy contained in large-scale structures increases with $\mathrm{Pm}$ proportionally to the total magnetic energy. Structures {at some of the largest allowed scales of the box} with $k<\pi/L$ and $k < 3\pi/L$ represent respectively about $10\%$ and $20\%$ of the total magnetic energy independently of $\mathrm{Pm}$. 

\section{Conclusion}
	\label{sec:conclusion}
We performed direct numerical simulations of MRI-driven dynamos with explicit viscosity and resistivity reaching unprecedentedly large values of the magnetic Prandtl number Pm. In the quasi-stationary state, the magnetic energy, kinetic energy and the angular momentum transport are approximately independent of the Reynolds number and follow a universal curve as a function of Pm. They first increase linearly with Pm up to moderately large values of this parameter (for $\mathrm{Pm}\lesssim50$), and smoothly transition to a regime consistent with a plateau independent of $\mathrm{Pm}$ at $\mathrm{Pm}\gtrsim100$. As $\mathrm{Pm}$ is increased, the peak of the magnetic energy shifts to larger wavenumbers. Interestingly, however, the energy contained in the largest scales of the magnetic field increases proportionally to the total magnetic energy, suggesting the presence of a large-scale dynamo whose efficiency increases with Pm. 
 
These results are particularly important for the formation of magnetars in fast rotating proto-neutron stars and in neutron star merger remnants. The increase with Pm of the magnetic energy contained at large scales would suggest that the dipolar component of the magnetic field should likewise increase in a spherical model. With this assumption, we may use our results to extrapolate the results of \citet{reboul-salze21} obtained at Pm=16 to the asymptotic regime of very high Pm relevant in a PNS. The fitting formula (\refeq{eq:fit}) predicts that the magnetic energy obtained by  \citet{reboul-salze21} is underestimated due to their moderate value of Pm by a factor $\mathrm{Pm_c}/16 \simeq 6$.   

In parallel to this work, \citet{held22} have also recently explored the MRI-driven dynamo in the regime of high magnetic Prandtl numbers. Several aspects of their results are qualitatively consistent with ours: the stress increases as a power-law with Pm at moderate Pm and transitions to a weaker dependence, possibly a plateau, at $\mathrm{Pm}> 50-100$. Some differences can nonetheless be noted and they may be explained by the different setup and parameter space exploration. First, they find a shallower power-law dependence  of the stress with Pm (power law index $0.5-0.7$) than the linear dependence that we find for moderate $\mathrm{Pm}$. This different slope is most likely primarily due to their choice of a Keplerian shear rate ($q=1.5$), while we studied a sub-Keplerian shear rate ($q=0.8$)\footnote{the three simulations performed by \citet{held22} at $q=0.8$ give a steeper Pm dependence with a power law index $\simeq1.4$.}. Their slope is consistent with previously published studies assuming Keplerian shear (see their Figure 17), although one should note that there is also a dependence on the box aspect ratio, such that \citet{simon09} for example is consistent with a linear dependence. The dependence on the shear rate and box aspect ratio should therefore be studied in more details in the future. {Another difference lies in the parameter space exploration: \citet{held22} ran series of simulations where Pm was varied at fixed Rm (by varying Re) while in our series of simulations Re was kept fixed and Rm varied. This might play a role in the other main difference with our results: while in our simulations the stress and magnetic energy are independent of the Reynolds number (for fixed Pm), \citet{held22} find an additional dependence on the magnetic Reynolds number at fixed Pm.} 

The dependence of the MRI saturated state on $\mathrm{Pm}$ (rather than $\mathrm{Re}$ or $\mathrm{Rm}$) in our simulations highlights the importance for numerical simulations to describe explicitly the diffusive processes. The effective $\mathrm{Pm}$ of implicit large eddy simulations \citep[e.g.][for the context of core collapse supernovae and neutron star mergers]{moesta15,kiuchi18} is of order unity regardless of the resolution they may reach. They are therefore very far from the high-$\mathrm{Pm}$ regime described in this paper and our results suggest that they may underestimate the magnetic energy by a factor up to 100.

Beyond the clear increase of the turbulent energies and stress with Pm, our results provide the first evidence that an asymptotic regime independent of Pm may exist for $\mathrm{Pm} \gtrsim 100$ (at least for the stress and kinetic energy). Such a regime had not been obtained by previously published MRI simulations, because they were restricted to relatively low Pm values ($\mathrm{Pm}\lesssim 16$). We propose that the plateau at very high Pm probably originates from the scale separation between the velocity and resistive scales, which prevents an efficient feedback of the resistive scales on the much larger scales of the flow. One should however caution that our simulations did not reach a very large scale separation between velocity and resistive scales, such that a weak dependence of the magnetic energy at high $\mathrm{Pm}$ cannot be excluded with the present data.  An extrapolation of our results to the much higher values of Pm relevant to proto-neutron stars is therefore still uncertain and will require a deeper analysis and physical understanding. In this perspective, it will be important to provide a robust physical explanation of the $\mathrm{Pm}$ dependence \citep{riols17,mamatshashvili20} and the asymptotic high-$\mathrm{Pm}$ regime with a detailed analysis of the energy transfers between different scales.

\section*{Acknowledgements}
JG, RR, and MB acknowledge support from the European Research Council
(MagBURST grant 715368). Numerical simulations have been carried out at the CINES on the Occigen supercomputer (DARI projects A0070410317, A0090410317 and A0110410317).

\section*{Data availability}
The data underlying this article will be shared on reasonable request to the corresponding author.

\bibliographystyle{mn2e}
\bibliography{supernovae}

\bsp
\label{lastpage}

\end{document}